\documentclass[prd,a4paper,twocolumn]{revtex4}
\usepackage{amsmath} 

\usepackage[T1] {fontenc}
\usepackage[latin1] {inputenc}

\usepackage{graphicx}
\usepackage[]{subfigure}

\newcommand{\diff}{ %differensial d
\mathrm{d}
}

\newcommand{\pderivop}[1]{ %lager partiellderivasjonsoperator
\frac{\partial}{\partial #1}
}

\newcommand{\deriv}[2]{ %lager totalderivert
\frac{\mathrm{d}#1}{\mathrm{d}#2}
}

\newcommand{\pderiv}[2]{ %lager partiellderivert
\frac{\partial#1}{\partial#2}
}

\newcommand{\bee}{\begin{equation}}
\newcommand{\ene}{\end{equation}}

\newcommand{\bea}{\begin{eqnarray}}
\newcommand{\eea}{\end{eqnarray}}
\newcommand{\bdm}{\begin{displaymath}}
\newcommand{\edm}{\end{displaymath}}

\begin{document}

\title{The Weyl curvature conjecture and black hole entropy}
\author{Øystein Rudjord}
\email{oysteir@ulrik.uio.no}
\affiliation{Department of physics, University of Oslo, Box 1048 Blindern, 0316 Oslo, Norway}
\author{Øyvind Grøn}
\email{oyvind.gron@iu.hio.no}
\affiliation{Department of physics, University of Oslo, Box 1048 Blindern, 0316 Oslo, Norway}
\affiliation{Oslo College, Faculty of engineering, Cort Adelers gt. 30, 0254 Oslo, Norway}

\date{\today}

%%%%%%%%%%%%%%%%%%%%%%%%%%%%%%%%%%%%%%%%%%%%%%%%%%%%%%%%%%%%%%%%%%%%%%%%%%
%%%%%%%%%%%%%%%%%%%%%%%%% Abstract %%%%%%%%%%%%%%%%%%%%%%%%%%%%%%%%%%%%%%%
%%%%%%%%%%%%%%%%%%%%%%%%%%%%%%%%%%%%%%%%%%%%%%%%%%%%%%%%%%%%%%%%%%%%%%%%%%

\begin{abstract}
The universe today, with structure such as stars, galaxies and black holes, seems 
to have evolved from a very homogeneous initial state.
From this it appears as if the entropy of the universe is decreasing, in violation 
of the second law of thermodynamics. 
It has been suggested by Roger Penrose \cite{grossmann:penrose:wcc} that this 
inconsistency can be solved 
if one assigns an entropy to the spacetime geometry. He also pointed out that
the Weyl tensor has the properties one would expect to find in a description of 
a gravitational entropy. In this article we make an attempt to use this 
so-called Weyl curvature conjecture to describe the Hawking-Bekenstein entropy
of black holes and the entropy of horizons due to a cosmological constant.
\end{abstract}
\maketitle

%%%%%%%%%%%%%%%%%%%%%%%%%%%%%%%%%%%%%%%%%%%%%%%%%%%%%%%%%%%%%%%%%%%%%%%%%%
%%%%%%%%%%%%%%%%%% Section 1: Gravitational entropy %%%%%%%%%%%%%%%%%%%%%%
%%%%%%%%%%%%%%%%%%%%%%%%%%%%%%%%%%%%%%%%%%%%%%%%%%%%%%%%%%%%%%%%%%%%%%%%%%

\section{Gravitational entropy}   
The second law of thermodynamics is well known as one of the most fundamental
laws of natural sciences. 
It captures most of our understanding of how macroscopic 
systems evolve over time. It can be stated as \emph{the entropy of a closed system 
never decreases}, or in a more mathematical form:
\begin{equation}
\deriv{S}{t} \ge 0.
\label{eq:second_law}
\end{equation}
This law introduces the concept of entropy, $S$. When it was introduced in the 
mid-nineteenth century it was initially considered as an
abstract thermodynamical quantity, but Ludwig Boltzmann later provided it with a 
physical interpretation. The entropy of a macroscopic state can be 
expressed from the multiplicity as
\begin{equation}
S = k_B \ln W
\label{eq:boltzmann_entropy}
\end{equation}
where $k_B=1.381\cdot 10^{-23}\mathrm J / \mathrm K$ is Boltzmann's constant
and the multiplicity, $W$, is the number of different configurations of 
microstates that results in the given macrostate. 
Together with (\ref{eq:second_law}) this means that a closed macroscopic system
will evolve towards the physically allowed state that can be represented by the 
largest possible number of microstates. This is a state of maximum 
entropy. 

For systems of everyday scales, this works just fine. A common example is 
that of an ideal gas trapped in a bottle. Once the bottle is opened, the gas will 
spread out and very soon fill all the available space. A state of 
maximum entropy has been reached. 

A gas of cosmological proportions, on the other hand, does not fit very well to 
this picture. Today most cosmologists agree that the universe started out in a 
very homogeneous state. Later gravitation made small density perturbations 
grow and ultimately formed structures such as stars, galaxies and black holes.  
This evolution appears to be the opposite of what happened in the previous example. 
Instead of spreading out, the gas collects itself into clumps of matter.

It should be noted, however, that linear perturbation theory shows that to first 
order in density- and corresponding temperature perturbations the thermodynamic 
entropy increases due to a transition of gravitational energy to thermal energy
\cite{gron_morad:entropy}, but this has not been shown for non-linear evolution 
from a nearly homogeneous initial state to a final inhomogeneous state 
with stars and galaxies. A gravity dominated evolution may represent a violation 
of the second law of thermodynamics as long as only thermodynamic entropy is
taken into account.

It was suggested by Roger Penrose \cite{grossmann:penrose:wcc} that this 
problem could be solved by assigning entropy to the gravitational field
itself. This means that a macroscopic state of spacetime geometry can be 
represented by many gravitational microstates. Without a working theory for
quantum gravitation, it is difficult to find a measure for this entropy. 
We will have to base our investigations on the behaviour of macroscopic systems.

Penrose \cite{grossmann:penrose:wcc} also suggested that the Weyl curvature tensor
could be used to 
quantify the gravitational entropy. 
The Weyl tensor is a rank four tensor that  
contains the independent components of the Riemann tensor  
not captured by the Ricci tensor.
It can be considered as the traceless part of the Riemann tensor, and
in four dimensions it can be expressed as
\[
C_{\alpha\beta\gamma\delta} = R_{\alpha\beta\gamma\delta}-g_{\alpha[\gamma}
R_{\delta]\beta} + g_{\beta[\gamma} R_{\delta]\alpha} + \frac 13 R 
g_{\alpha[\gamma}g_{\delta]\beta}.
\]
where the bracketed indices should be understood as antisymmetric
combinations ($A_{[\mu\nu]} = \frac 12 \left(A_{\mu\nu}-A_{\nu\mu}\right)$).

The Weyl tensor is zero
in the Friedmann-Robertson-Walker models, which describe the 
early universe quite well on small scales as well as on large scales since the 
universe was nearly homogeneous on all macroscopic scales before the first 
stars formed. It is also large in the Schwarzschild
spacetime, which describes the spacetime geometry around a spherically symmetric 
mass distribution such as a black hole and other phenomena of the late universe. 
This is the kind of behaviour we would expect 
from a description of gravitational entropy, since it would be increasing 
throughout the history of the universe, in agreement with the second law. 

Penrose \cite{grossmann:penrose:wcc} conjectured that the evolution of the 
universe started from an initial 
singularity in which the Weyl tensor was much less than the Ricci part of the 
spacetime curvature. Ø. Grøn and S. Hervik \cite{gron+hervik:weyl2} have later 
used a
mini-superspace model to perform a quantum mechanical calculation of the 
probability that this so-called Weyl curvature conjecture is true.
The result supported the conjecture. Also, J. D. Barrow and S. Hervik 
\cite{barrow+hervik:weyl}
have studied the evolution of the Weyl curvature invariant in isotropic, 
homogeneous universe models, and shown that the Weyl curvature invariant 
dominates the Ricci invariant at late times. 

The Weyl curvature tensor is also locally independent of the energy-momentum 
tensor, which makes the hypothesis more convincing. If it were coupled to the 
matter and energy content of spacetime (like the Ricci curvature), 
we would have to account for a degeneracy 
between gravitational microstates and the common, ``thermodynamical'' microstates
of the matter.

There is of course a question whether the Weyl curvature tensor can be considered
as a direct measure of gravitational entropy or if its
entropy-like behaviour is only an effect of another process more closely related 
to the entropy. 
Nevertheless, it seems to be a natural place to start our investigations if we 
seek to understand gravitational entropy.

First, in section \ref{sec:entropy_of_black_holes} we will briefly review 
black holes and the entropy associated with them.
In section \ref{sec:black_holes_end_weyl_entropy} we investigate possibilities 
for defining an expression involving the Weyl curvature scalar which can be used 
in the construction of a quantity representing gravitational entropy. A
guideline for this construction is that it shall provide a geometrical 
interpretation of the Hawking-Bekenstein entropy.
In section \ref{sec:sds} we add a cosmological constant
to the system and apply the model for gravitational entropy to the 
Schwarzschild-de~Sitter spacetime. Finally, in section \ref{desitter}
we remove the black hole and take a look at the de~Sitter spacetime.

%%%%%%%%%%%%%%%%%%%%%%%%%%%%%%%%%%%%%%%%%%%%%%%%%%%%%%%%%%%%%%%%%%%%%%%%%%
%%%%%%%%%%%%%%%%%% Section 2: Entropy of black holes %%%%%%%%%%%%%%%%%%%%%
%%%%%%%%%%%%%%%%%%%%%%%%%%%%%%%%%%%%%%%%%%%%%%%%%%%%%%%%%%%%%%%%%%%%%%%%%%

\section{Entropy of black holes}\label{sec:entropy_of_black_holes}
The \emph{no hair theorem} \cite{misner+thorne+wheeler:gravitation} states that
a black hole is completely described by three characteristics: mass, charge
and angular momentum. Classically, all other information is lost when 
something falls beyond the event horizon. Therefore, if one drops a package of
entropy into a black hole, the entropy seems to disappear from the universe
\footnote{This was first pointed out by J. A. Wheeler, as mentioned 
by J. Bekenstein in \cite{bekenstein:black_hole_entropy}.}. 
Again, the second law appears to be violated. 

However, Hawking and Christodoulou \cite{hawking:area_theorem}
\cite{hawking:black_holes_in_gr} \cite{christodoulou:area}
\cite{christodoulou+ruffini:area} showed that the horizon area of a black hole
(or indeed the total area of several black holes) never decreases, and is only 
constant for a special class of transformations. The parallel to the second law 
seems obvious, but it was Bekenstein \cite{bekenstein:black_hole_second_law} 
\cite{bekenstein:black_hole_entropy} \cite{bekenstein:generalized_second_law}
who made the bold proposition that the
horizon area actually was a measure of the entropy of the black hole.
This entropy, called the Hawking-Bekenstein entropy is expressed 
as
\begin{equation}
S_{HB} = \frac{k_B}{4l_P^2}A = \frac{k_Bc^3}{4G\hbar}A
\label{eq:hb_entropy}
\end{equation}
where $A$ is the area of the black hole horizon, $l_P^2 =\frac{G\hbar}{c^3}$ is
the Planck area, $G = 6.673\cdot 10^{-11}\mathrm{Nm}^2/\mathrm{kg}^2$ is the 
constant of gravitation and $\hbar = \frac{h}{2\pi}= 1.055\cdot 10^{-34}\mathrm{Js}$
where $h$ is Planck's constant.

The total entropy is then the sum of the Hawking-Bekenstein entropy and the common
thermodynamical entropy:
\[
S_{tot} = S_{td} + S_{HB}
\]
and this entropy never decreases, in agreement with the second law.

%%%%%%%%%%%%%%%%%%%%%%%%%%%%%%%%%%%%%%%%%%%%%%%%%%%%%%%%%%%%%%%%%%%%%%%%%%
%%%%%%%%%%%%%%%% Section 3: Black holes and Weyl entropy %%%%%%%%%%%%%%%%%
%%%%%%%%%%%%%%%%%%%%%%%%%%%%%%%%%%%%%%%%%%%%%%%%%%%%%%%%%%%%%%%%%%%%%%%%%%

\section{Black holes and Weyl entropy}\label{sec:black_holes_end_weyl_entropy}
We will now assume that the Hawking-Bekenstein entropy is of geometrical origin, 
and is therefore a special case of gravitational entropy. The total
entropy of a system will then be the sum of thermodynamical entropy and 
gravitational entropy
\[
S_{tot}=S_{td}+S_{grav}.
\]
We will use this 
assumption to develop a more general model for gravitational entropy
for spherically symmetric spacetimes. 

%%%%%%% Subsection A: Construction of an expression for gravitational entropy 

\subsection{Construction of an expression for gravitational entropy}
First we need to make a more general description of the Hawking-Bekenstein entropy.
It is proportional to the horizon area, so we assume that it can be expressed 
as a surface integral of some vector field $\vec \Psi$ over the horizon:
\begin{equation}
S = k_S \int_\sigma \vec \Psi \cdot \vec{\diff\sigma}
\label{eq:entropy_surf_int}
\end{equation}
where $k_S$ is an unknown constant. Our goal now, is to find a vector field, $\vec 
\Psi$ that gives a convincing description of entropy for a black hole. Then
we demand that this is equal to the Hawking-Bekenstein entropy
\[
S = S_{HB}
\]
and use this to determine $k_S$.

By means of Gauss' divergence theorem, we can rewrite 
the surface integral (\ref{eq:entropy_surf_int}) as a volume integral:
\[
\int_\sigma \vec \Psi \cdot \vec{d\sigma} = \int_V \left(\nabla \cdot \vec \Psi 
\right)\diff V.
\]
It is then convenient to define an entropy density:
\begin{equation}
s = k_S \nabla \cdot \vec \Psi.
\label{eq:def_entropy_density}
\end{equation}

Now, we assume that the vector field is radial:
\begin{equation}
\vec \Psi = P\vec e_{\hat r}.
\label{eq:def_psi}
\end{equation}
Since we intend to describe the entropy according to the 
Weyl curvature conjecture, we assume that $P$ is some scalar derived from the Weyl 
tensor \cite{gron+hervik:weyl}. At this point we have no further clue about how $P$ should be found, 
so we start by trying some of the easiest ways to extract a scalar from the 
Weyl tensor.

%%%%%%%%%%%%%% Subsection B: The Weyl scalar %%%%%%%%%%%%%%%%%%%%%%%%%%%%%%%%

\subsection{The Weyl scalar}
Usually, the easiest way to get a scalar value from a tensor is 
find the trace, by contracting the indices. The Weyl tensor however, is 
traceless, so this always yields zero.
\[
C^{\alpha\beta}_{\ \ \alpha \beta} = 0 
\]

The next easiest way to create a scalar is to ``square'' it. We therefore 
start by defining
\begin{equation}
P^2 = C^{\alpha\beta\gamma\delta}C_{\alpha\beta\gamma\delta}.
\label{eq:def_p_weylscalar}
\end{equation}
We will refer to this quantity as the Weyl scalar.

For now, we will constrain ourselves to non-rotating electrically neutral 
black holes, 
and therefore we can describe the spacetime geometry with the Schwarzschild
metric:
\begin{equation}
\diff s^2 = -e^{\nu(r)}c^2 \diff t^2
+e^{\lambda(r)} \diff r^2 + r^2\diff \theta^2 +
r^2\sin^2\theta \diff \phi^2
\label{eq:Schwarzschild_metric}
\end{equation}
where 
\begin{equation}
e^{\nu(r)} = e^{-\lambda(r)} = 1-\frac {R_S}{r}. 
\label{eq:nu_and_lambda}
\end{equation}
and $R_S = \frac{2MG}{c^2}$ is the Schwarzschild radius.

The nonzero, independent components of the Weyl tensor in this spacetime
is given as \cite{leon:weyl}
\begin{equation}
\begin{split}
C^{\hat 3}_{\hat 0\hat 3\hat 0} &= C^{\hat 2}_{\hat 0\hat 2\hat 0}
=-C^{\hat 1}_{\hat 3\hat 1\hat 3}=-C^{\hat 1}_{\hat 2\hat 1\hat 2} \\
&=\frac 12 C^{\hat 3}_{\hat 2\hat 3\hat 2}
=-\frac 12 C^{\hat 1}_{\hat 0\hat 1\hat 0}= \frac{e^{-\lambda(r)}}{6}\alpha(r)
\end{split} \label{eq:weyl_indep_comp}
\end{equation}
where $\alpha(r)$ is 
\begin{equation}
\begin{split}
\alpha(r) &= \frac{e^\lambda(r)}{r^2}-\frac{1}{r^2}-\frac{\nu'(r)^2}{4} \\
&+\frac{\nu'(r)\lambda'(r)}{4} -\frac{\nu''(r)}{2}-\frac{\lambda'(r)-\nu'(r)}{2r}.
\end{split}
\label{eq:alpha}
\end{equation}

The Weyl tensor inherits the symmetries of the Riemann tensor:
\begin{eqnarray}
C_{\alpha\beta\gamma\delta}&=& C_{\gamma\delta\alpha\beta} \\
C_{\alpha\beta\gamma\delta}&=& -C_{\beta\alpha\gamma\delta} \\
C_{\alpha\beta\gamma\delta}&+&C_{\alpha\delta\beta\gamma}
+C_{\alpha\gamma\delta\beta} = 0. \label{eq:weyl_sym2}
\end{eqnarray}
Using eqs. (\ref{eq:weyl_indep_comp})-(\ref{eq:weyl_sym2}) in the calculation 
of the Weyl scalar, we find
\begin{equation}
C^{\alpha\beta\gamma\delta}C_{\alpha\beta\gamma\delta} =
12\left(\frac{e^{\lambda(r)}}{3}\alpha(r)\right)^2.
\label{eq:weyl_scalar_general}
\end{equation}
For the Schwarzschild metric we use (\ref{eq:nu_and_lambda}) and 
(\ref{eq:alpha}) to find 
\begin{equation}
\alpha(r) = \frac{3R_S}{r^3}\frac{1}{1-\frac{R_S}{r}},
\label{eq:alpha_schw}
\end{equation}
resulting in the Weyl scalar in the Schwarzschild spacetime
\begin{equation}
C^{\alpha\beta\gamma\delta}C_{\alpha\beta\gamma\delta} =
12\left(\frac{R_S}{r^3}\right)^2 = \frac{48M^2G^2}{r^6c^4}
\label{eq:weyl_scalar_schw}
\end{equation}

Combining the Weyl scalar with (\ref{eq:def_p_weylscalar}) we find (assuming 
the positive solution) that
\begin{equation}
P = 2\sqrt 3 \frac{R_S}{r^3}.
\label{eq:p_weylscalar_schw}
\end{equation}
We then use this with (\ref{eq:entropy_surf_int}) to find the entropy.
Since we are operating in $3$-space, we must at this point find a spatial 
metric. We define the spatial metric $h_{ij}$ as
\[
h_{ij} = g_{ij}-\frac{g_{i0}g_{j0}}{g_{00}}
\]
where $g_{\mu\nu}$ is the $4$-dimensional spacetime metric and Latin indices denote
spatial components, $i,j = 1,2,3$.
This yields 
\[
h_{ij} = \mathrm{diag}(e^{\lambda(r)}, r^2, r^2\sin\theta)
\]
The infinitesimal surface element on the horizon surface is then
\begin{eqnarray}
\vec{\diff \sigma} &=& \vec e_{\hat{r}} \frac{\sqrt{h}}{\sqrt {h_{rr}}}\diff\theta\diff\phi \notag \\
&=& \vec e_{\hat{r}} r^2 \sin\theta \diff\theta\diff\phi
\label{eq:surf_element}
\end{eqnarray}

Then we use (\ref{eq:entropy_surf_int}), (\ref{eq:p_weylscalar_schw}), 
(\ref{eq:def_psi}) and (\ref{eq:surf_element}) to find the entropy.
Since there is a singularity at the origin, we must be careful about integrating 
around it. Therefore we also integrate over a small sphere
with radius $\varepsilon$ around the origin, and subtract this from the integral
(\ref{eq:entropy_surf_int}). When we let $\varepsilon \rightarrow 0$ this should 
result in the entropy. The result is
\begin{eqnarray}
S &=& k_{S1} \left( P(R_S)R_S^2 -P(\varepsilon)\varepsilon^2
\right) \int_0^{2\pi} \int_0^\pi \sin \theta \diff \theta \diff \phi \notag \\
&=& 4\pi k_{S1} \left(2\sqrt 3 -2\sqrt 3 \frac{R_S}{\varepsilon}\right).
\end{eqnarray}
Here, we see two things. First, when we let $\varepsilon \rightarrow 0$
this entropy diverges. Secondly, this result is not proportional to the area.
We should therefore find a different scalar, $P$, from the Weyl tensor.

%%%%%%%%%%%%%% Subsection C: The surface gravity %%%%%%%%%%%%%%%%%%%%%%%%%%%%%%%%

\subsection{The surface gravity}
Let us now redefine 
\begin{equation}
P^2 = \frac{C^{\alpha\beta\gamma\delta}C_{\alpha\beta\gamma\delta}}{\kappa^4(r)}
\label{eq:def_p_kappa}
\end{equation}
where $\kappa(r)$ is the acceleration of gravity. In the Schwarzschild spacetime 
this is \cite{gron:repulsive} 
\begin{equation}
\kappa(r) = -\frac{R_Sc^2}{2r^2}
\label{eq:kappa_schw}
\end{equation}
From (\ref{eq:def_p_kappa}), (\ref{eq:weyl_scalar_schw}) and (\ref{eq:kappa_schw})
we find
\begin{equation}
P = \frac{8\sqrt 3}{R_S c^4}r
\label{eq:p_kappa_schw}
\end{equation}

Again, we calculate the entropy (\ref{eq:entropy_surf_int}), 
(\ref{eq:p_kappa_schw}), (\ref{eq:def_psi}) and (\ref{eq:surf_element})
using the same approach as in the previous section
\begin{eqnarray}
S &=& k_{S2} \left( P(R_S)R_S^2 -P(\varepsilon)\varepsilon^2
\right) \int_0^{2\pi} \int_0^\pi \sin \theta \diff \theta \diff \phi \notag \\
&=& 4\pi k_{S2} \left(\frac{8\sqrt 3}{c^4}R_S^2 -\frac{8\sqrt 3}{R_S c^4}
\varepsilon^3\right). 
\end{eqnarray}
If we let $\varepsilon \rightarrow 0$ here, the entropy of a black hole becomes
\begin{equation}
S =  k_{S2} \frac{32\sqrt 3\pi}{c^4}R_S^2 = k_{S2} \frac{8\sqrt 3}{c^4}A
\label{eq:entropy_schw_kappa}
\end{equation}
where $A = 4\pi R_S^2$ is the horizon area of a black hole. 
Comparing with the Hawking-Bekenstein entropy (\ref{eq:hb_entropy}) and demanding
$S = S_{HB}$ yields
\[
k_{S2} = \frac{k_Bc^7}{32\sqrt 3 G \hbar}.
\]

Note that the entropy is proportional to the area only for 
the case of a black hole horizon. The gravitational entropy
of a general spherical mass distribution with radius $R$ yields
\begin{equation}
S =  k_{S2} \frac{32\sqrt 3\pi R^3}{R_S c^4}= k_{S2} \frac{24 \sqrt 3}{R_Sc^4}V
\label{eq:entropy_schw_kappa_volume}
\end{equation}
where $V = \frac 43 \pi R^3$ is the volume of the sphere. 

Recall that we used Gauss' integral theorem to define entropy density 
in (\ref{eq:def_entropy_density}). Calculating the divergence of 
$\vec \Psi$ we find
\begin{equation}
s = k_{S2} \nabla \cdot \vec \Psi = k_{S2} \frac{24 \sqrt 3}{R_Sc^4}
\sqrt{1-\frac{R_S}{r}} 
\label{eq:entropy_density_schw_kappa}
\end{equation}
We see that this quantity becomes imaginary inside the horizon ($r < R_S$)
even if the entropy (\ref{eq:entropy_schw_kappa}) is real.
If we instead use (\ref{eq:entropy_schw_kappa_volume}) to define 
a ``physical'' entropy density, $\hat s$, by simply factoring out the volume, $V$,
we find 
\begin{equation}
\hat s =k_{S2} \frac{24 \sqrt 3}{R_Sc^4}
\label{eq:phys_entropy_density_kappa}
\end{equation}

The physical entropy density (\ref{eq:phys_entropy_density_kappa}) is 
constant for a given mass of a black hole, and the ``mathematical'' entropy
density (\ref{eq:entropy_density_schw_kappa}) approaches a constant positive 
value for $r \gg R_S$.
As the spacetime is asymptotically Minkowski far away from the black hole, 
we would expect the entropy density to vanish for $r 
\rightarrow \infty$. What we have found however, is that the gravitational
entropy is non-vanishing even in the Minkowski spacetime. Since this seems 
quite unphysical, we should look for another expression of gravitational entropy.

%%%%%%%%%%%%%% Subsection D: The Ricci tensor %%%%%%%%%%%%%%%%%%%%%%%%%%%%%%%%

\subsection{The Ricci tensor}
A quantity that has been much discussed in relation to the Weyl curvature conjecture
is the ratio between the Weyl scalar and the squared Ricci tensor. The Ricci
tensor seems to have opposite properties to the Weyl tensor with respect to 
gravitational entropy, and therefore this ratio has been suggested as an alternative
to using only the Weyl tensor. It has also been pointed out that this quantity
may be more well behaved in an initial 
singularity \cite{goode+wainwright:singularities} \cite{wainwright:singularities}.
Therefore we redefine $P$ again: 
\begin{equation}
P^2 = \frac{C^{\alpha\beta\gamma\delta}C_{\alpha\beta\gamma\delta}}{R^{\mu\nu}
R_{\mu\nu}}.
\label{eq:def_p_ricci}
\end{equation}

The first problem we encounter is that the Ricci tensor is zero in vacuum and 
therefore also in the Schwarzschild spacetime. In an attempt to solve this, we 
invoke the Hawking radiation \cite{hawking:radiation}. One may hope that the 
radiation from the horizon will curve the spacetime slightly and prevent 
the Ricci tensor from vanishing. In order to describe this, we use the Vaidya 
spacetime
\begin{equation}
\diff s^2 = -\left(1-\frac{2MG}{rc^2}\right)\diff v^2 +2\diff v\diff r
+r^2\diff \theta^2 +r^2\sin^2\theta \diff \phi^2
\end{equation}
where $v=t+r^*$ is an advanced time coordinate, $r^*= 
r +\frac{2MG}{c^2}\ln\left(\frac{rc^2}{2MG}-1\right)$ and $M = M(v)$.

The Weyl scalar in this spacetime is \cite{barve_singh:weyl}
\begin{equation}
C^{\alpha\beta\gamma\delta}C_{\alpha\beta\gamma\delta}
=\frac{48M^2(v)G}{r^6c^4}.
\end{equation}
To find the squared Ricci tensor we start with Einstein's field equations
\begin{equation}
R_{\mu\nu} -\frac 12 R g_{\mu\nu} = \frac{8\pi G}{c^4}T_{\mu\nu}
\end{equation}
this gives
\[
R^{\mu\nu}R_{\mu\nu} = \left(\frac{8\pi G}{c^4}\right)^2T^{\mu\nu}T_{\mu\nu}
\]
The only nonzero covariant component of the energy momentum tensor for 
radiation is \cite{barve_singh:weyl}
\begin{equation}
T_{vv}= \frac{2c^2}{8\pi r^2}\pderiv{M}{v}
\end{equation}
However, the only nonzero contravariant component of the energy momentum tensor 
is
\[
T^{rr}= \frac{2c^2}{8\pi r^2}\pderiv{M}{v}.
\]
Thus the product $T^{\mu\nu}T_{\mu\nu}$ is zero, and
\begin{equation}
R^{\mu\nu}R_{\mu\nu} = 0 
\end{equation}
for the Vaidya spacetime. Hence (\ref{eq:def_p_ricci}) is non-defined, and again
we must find another expression for $P$.

%%%%%%%%%%%%%% Subsection E: The Riemann tensor %%%%%%%%%%%%%%%%%%%%%%%%%%%%%%%%

\subsection{The Riemann tensor}\label{sec:riemann}
We will now investigate an expression for $P$ where we use the Kretschmann scalar
instead of the squared Ricci tensor:
\begin{equation}
P^2 = \frac{C^{\alpha\beta\gamma\delta}C_{\alpha\beta\gamma\delta}}
{R^{\alpha\beta\gamma\delta}R_{\alpha\beta\gamma\delta}}
\label{eq:def_p_kretschmann}
\end{equation}
For static spherically symmetric spacetimes, $0\le P^2\le 1$ (see appendix 
\ref{app}). This may possibly be connected to states of zero gravitational entropy
and maximum gravitational entropy respectively.

Since the Kretschmann scalar is not necessarily zero in vacuum, we return to the 
Schwarzschild spacetime (\ref{eq:Schwarzschild_metric})-(\ref{eq:nu_and_lambda}).
In this metric all curvature is Weyl curvature, so that 
\[
R^{\alpha\beta\gamma\delta}R_{\alpha\beta\gamma\delta}
=C^{\alpha\beta\gamma\delta}C_{\alpha\beta\gamma\delta}
=12\left(\frac{R_S}{r^3}\right)^2 
\]
and $P^2$ becomes simply
\[
P^2= 1.
\]
If our speculations are correct, this means that a Schwarzschild black hole
is a configuration of maximum gravitational entropy. 

We find the entropy of a black hole as before:
\begin{eqnarray}
S &=& k_{S3} \int_\sigma \vec\Psi \cdot \vec{\diff \sigma} \notag \\
&=& k_{S3} \int_\sigma \left(R_S^2 -\varepsilon^2 \right)\sin\theta 
\diff \theta \diff \phi \notag \\
&=& k_{S3} 4\pi \left(R_S^2 -\varepsilon^2\right).\notag 
\end{eqnarray}
Letting $\varepsilon \rightarrow 0$ we get
\begin{equation}
S = k_{S3} 4\pi R_S^2 = k_{S3}A
\label{eq:entropy_schw_kret}
\end{equation}
where $A = 4\pi R_S^2$ is the horizon area of the black hole. Comparing with
the Hawking-Bekenstein entropy and demanding $S = S_{HB}$ yields
\begin{equation}
k_{S3} = \frac{k_B}{4l_P^2}=\frac{k_Bc^3}{4G\hbar}
\label{eq:const_k_s3}
\end{equation}

Now, we proceed to find the entropy density in the Schwarzschild spacetime.
From the definition of entropy density (\ref{eq:def_entropy_density}) 
we find
\begin{eqnarray}
s &=& k_{S3} \nabla \cdot \vec \Psi  \notag \\
&=& k_{S3} \frac{1}{\sqrt{h}}\pderivop{r}\left(\sqrt h \frac{P}{\sqrt{h_{rr}}}
\right) \notag \\
&=& \frac{2k_{S3}}{r} \sqrt{1-\frac{R_S}{r}}
\label{eq:entropy_density_schw}
\end{eqnarray}
where we have used the covariant divergence, $\nabla \cdot \vec \Psi =
\frac{1}{\sqrt{h}}\pderivop{x^\mu}\left(\sqrt h \Psi^\mu\right)$.

This entropy density vanishes as $r \rightarrow \infty$ and the spacetime becomes 
Minkowski. The entropy density reaches a maximum at $r = \frac 32 R_S$, and then 
vanishes at the horizon. Inside the horizon, it becomes imaginary.

It should be noted, that for such a generalization, 
the entropy of the entire space diverges as $A\rightarrow \infty$ in 
eq. (\ref{eq:entropy_schw_kret}).

%%%%%%%%%%%%%%%%%%%%%%%%%%%%%%%%%%%%%%%%%%%%%%%%%%%%%%%%%%%%%%%%%%%%%%%%%%
%%%%%%%%%%%%% Section 4: The Schwarzschild-de~Sitter spacetime %%%%%%%%%%%
%%%%%%%%%%%%%%%%%%%%%%%%%%%%%%%%%%%%%%%%%%%%%%%%%%%%%%%%%%%%%%%%%%%%%%%%%%

\section{The Schwarzschild-de~Sitter spacetime}\label{sec:sds}
We will now see how this description of gravitational entropy applies to a black 
hole in de~Sitter background. If we include a cosmological constant into the 
field equations
\[
R_{\mu\nu}-\frac 12 Rg_{\mu\nu} + \Lambda g_{\mu\nu} = \frac{8\pi G}{c^4}T_{\mu\nu}
\]
and solve for a spherically symmetric mass distribution 
we get the Schwarzschild-de~Sitter (SdS) spacetime. This metric can be written on 
the form
(\ref{eq:Schwarzschild_metric}) where 
\begin{equation}
e^{\nu(r)}= e^{-\lambda(r)} = 1-\frac{\Lambda r^2}{3} - \frac{R_S}{r}.
\end{equation}

The Weyl scalar in the SdS spacetime is the same as in the Schwarzschild
spacetime
\begin{equation}
C^{\alpha\beta\gamma\delta}C_{\alpha\beta\gamma\delta}
=\frac{48M^2G^2}{r^6c^4}
\end{equation}
and the Kretschmann scalar is
\begin{equation}
R^{\alpha\beta\gamma\delta}R_{\alpha\beta\gamma\delta} 
= \frac{8(18M^2G^2+\Lambda^2 r^6c^4)}{3r^6c^4}.
\end{equation}

We apply the expression (\ref{eq:def_p_kretschmann}) for $P$ from the 
previous section and find
\[
P = \frac{1}{\sqrt{1+\frac{\Lambda^2c^4}{18M^2G^2}r^6}}.
\]
Together with (\ref{eq:entropy_surf_int}) we use this to find the 
entropy of the black hole at the center of this spacetime as before.
\begin{eqnarray}
S &=& k_{S3} \int_\sigma \vec\Psi \cdot \vec{\diff \sigma} \notag \\
&=& k_{S3} \int_\sigma \left(P(R_h)R_h^2 -P(\varepsilon)\varepsilon^2 \right)
\sin\theta \diff \theta \diff \phi \notag \\
&=& k_{S3} 4\pi \left(P(R_h)R_h^2 - P(\varepsilon) \varepsilon^2\right)\notag \\
&=& k_{S3} 4\pi \left[ \frac{R_h^2}{\sqrt{1+\frac{\Lambda^2 c^4}{18M^2G^2}
R_h^6}} - \frac{\varepsilon^2}{\sqrt{1+\frac{\Lambda^2 c^4}{18M^2G^2}
\varepsilon^6}} \right] \notag 
\end{eqnarray}
where 
\begin{equation}
R_h = -\frac{2}{\sqrt \Lambda} \cos \left[\frac{\arccos \left(-\frac{3MG\sqrt\Lambda}{c^2}\right)+\pi}{3}\right]
\end{equation}
is the radius of the black hole horizon and $k_{S3}$ can be found 
from (\ref{eq:const_k_s3}). We let $\varepsilon \rightarrow 0$ and get the 
entropy of the black hole
\begin{equation}
S =  \frac{k_{S3} 4\pi R_h^2}{\sqrt{1+\frac{\Lambda^2 c^4}{18M^2G^2}
R_h^6}} 
\end{equation}
From this we see that the entropy decreases for larger values of $\Lambda$, 
and if we let $\Lambda \rightarrow 0$ the black hole entropy becomes the 
Hawking-Bekenstein entropy, $S = k_{S3}A$, where $A= 4\pi R_h^2$ is the 
black hole horizon area.

\begin{figure}
 \begin{center}
  \includegraphics[height=8cm, width=8cm]{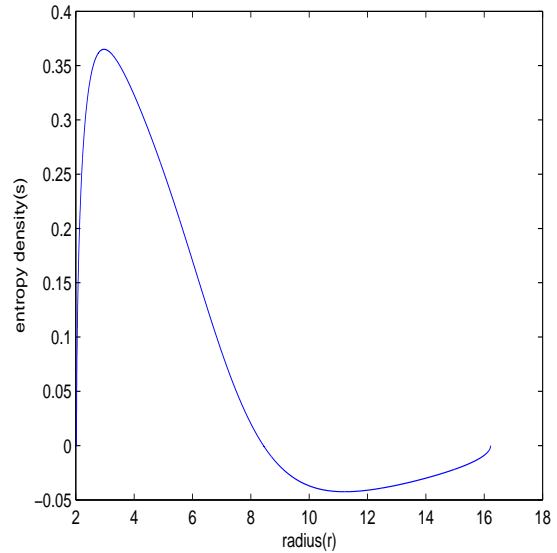}
 \end{center}
\caption{The entropy density of the Schwarzschild-de~Sitter spacetime between 
the horizons. Here, $c= G =k_{S3} = 1$, $M=1$ and $\Lambda=0.01$.}
\label{fig:sds_entropy_density}
\end{figure}

The entropy density in this spacetime can be found in the same way as 
in the previous section (\ref{sec:riemann}), using the definition
(\ref{eq:def_entropy_density}), and it becomes 
\begin{equation}
s = \frac{k_{S3}}{r}
\sqrt{1-\frac{\Lambda r^2}{3} - \frac{R_S}{r}}\left(
\frac{2-\frac{\Lambda^2c^4}{18M^2G^2}r^6}{\left(1+\frac{\Lambda^2c^4}{18M^2G^2}r^6 
\right)^{3/2}}\right).
\label{eq:entropy_density_sds}
\end{equation}
Also here we see that if $\Lambda \rightarrow 0$ this reduces to the 
same result as in the Schwarzschild spacetime (\ref{eq:entropy_density_schw}).
The density (\ref{eq:entropy_density_sds}) is plotted in 
figure \ref{fig:sds_entropy_density}.
Obviously this shows a region with negative entropy density. 
This seems quite unphysical, especially when comparing with
Boltzmann's interpretation of entropy (\ref{eq:boltzmann_entropy}), then 
this would correspond to a multiplicity of less than one. 

To solve this problem we demand that the entropy density is positive.
\begin{equation}
s = k_{S3} \left| \nabla \cdot \vec \Psi \right|
\label{eq:e_dens_abs}
\end{equation}
This can be done by choosing the sign of $P$. It is always defined as a square, 
and then we choose the positive square root, $P = +\sqrt{P^2}$. Instead we could 
always choose the sign that results in positive entropy density. 

\begin{figure}
 \begin{center}
  \includegraphics[height=8cm, width=8cm]{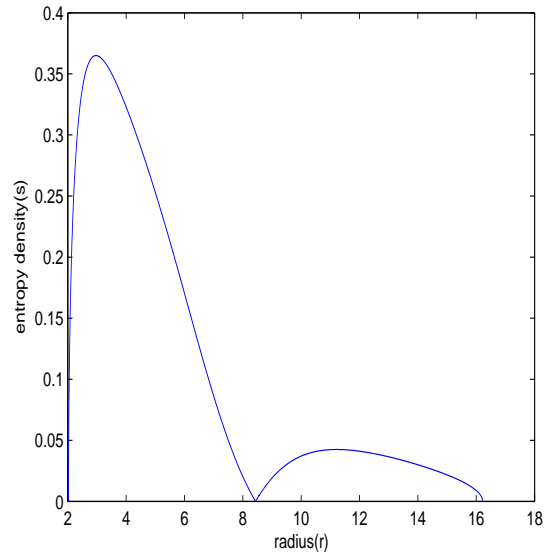}
 \end{center}
\caption{The entropy density of the Schwarzschild-de~Sitter spacetime 
between the horizons according to eq. (\ref{eq:e_dens_abs}). Here, 
$c= G =k_{S3} = 1$, $M=1$ and $\Lambda=0.01$.}
\label{fig:sds_entropy_density_abs}
\end{figure}

For the Schwarzschild-de~Sitter space, the entropy density is zero at
\begin{equation}
r_0 = \left(\frac{6MG}{\Lambda c^2}\right)^{1/3}.
\end{equation}
By demanding that it is positive also outside this radius, the entropy density
behaves as plotted in figure \ref{fig:sds_entropy_density_abs}.

%%%%%%%%%%%%%%%%%%%%%%%%%%%%%%%%%%%%%%%%%%%%%%%%%%%%%%%%%%%%%%%%%%%%%%%%%%
%%%%%%%%%%%%% Section 5: The de~Sitter spacetime %%%%%%%%%%%%%%%%%%%%%%%%%
%%%%%%%%%%%%%%%%%%%%%%%%%%%%%%%%%%%%%%%%%%%%%%%%%%%%%%%%%%%%%%%%%%%%%%%%%%

\section{The de~Sitter spacetime}\label{desitter}
The SdS spacetime has two obvious special cases. 
One is obtained by putting $\Lambda = 0$ resulting in Schwarzschild spacetime.
The other is known as the de~Sitter spacetime and is obtained by putting
$M=0$. The line element in this spacetime 
may also be expressed as in (\ref{eq:Schwarzschild_metric}), but with
\begin{equation}
e^{\nu(r)}= e^{-\lambda(r)} = 1-\frac{\Lambda r^2}{3}.
\end{equation}
This spacetime has no black hole, only a cosmological horizon at \mbox{$R_\Lambda = 
\sqrt{\frac{3}{\Lambda}}$.} 

G. Gibbons and S. W. Hawking \cite{hawking+gibbons:event_horisons_td}
found that this horizon has entropy similar to a black hole horizon,
\begin{equation}
S_\Lambda = \frac{k_B c^3}{4G\hbar}A.
\end{equation}
where $A = 4\pi R_\Lambda^2 = \frac{12\pi}{\Lambda}$ is the area of the 
cosmological horizon.

The Weyl curvature tensor vanishes in de~Sitter spacetime, and consequently 
the gravitational entropy in this spacetime is zero.
This conflicts with the idea of reconciling horizon entropy with gravitational 
entropy. This could of course be solved by developing some model where 
the gravitational entropy is non-vanishing in de~Sitter spacetime. It
would give little satisfaction however, since the de~Sitter spacetime 
gives a very good description of the early universe where the gravitational 
entropy is expected to be small (or zero). The fact that the Weyl tensor 
vanishes in this spacetime was one of the reasons for using it to describe 
gravitational entropy in the first place. 
Thus the gravitational entropy is expected to vanish in this spacetime, 
and therefore the entropy of the cosmological horizon must be of 
non-geometrical origin.

%%%%%%%%%%%%%%%%%%%%%%%%%%%%%%%%%%%%%%%%%%%%%%%%%%%%%%%%%%%%%%%%%%%%%%%%%%
%%%%%%%%%%%%% Section 6: Summary and conclusions %%%%%%%%%%%%%%%%%%%%%%%%%
%%%%%%%%%%%%%%%%%%%%%%%%%%%%%%%%%%%%%%%%%%%%%%%%%%%%%%%%%%%%%%%%%%%%%%%%%%

\section{Summary and conclusions}
We have now developed a possible description of gravitational entropy
motivated by the possibility of reconciling it with the Hawking-Bekenstein
entropy of black holes. We have assumed the entropy of a black hole to be
described by the surface integral 
\begin{equation}
S = k_S \int_\sigma \vec \Psi \cdot \vec{\diff \sigma}
\label{conc:entropy_int}
\end{equation}
where $\sigma$ is the horizon of the black hole and the vector field $\vec \Psi$
is 
\[
\vec \Psi = P\vec e_{\hat r}.
\]
Comparison with the Hawking-Bekenstein entropy led us to define $P^2$ as
the ratio of the Weyl scalar and the Kretschmann scalar:
\[
P^2 = \frac{C^{\alpha\beta\gamma\delta} C_{\alpha\beta\gamma\delta}}
{R^{\alpha\beta\gamma\delta} R_{\alpha\beta\gamma\delta}}.
\]
This resulted in 
\[
k_S = k_{S3} = \frac{k_B}{4l_P^2}=\frac{k_Bc^3}{4G\hbar}.
\]

By means of Gauss' divergence
theorem (\ref{conc:entropy_int}) can be rewritten as a volume integral, and 
with this in mind we defined the entropy density 
\[
s = k_S \left| \nabla \cdot \vec \Psi \right|
\]
where the absolute value brackets was added to avoid negative entropy.

After a brief investigation of the de~Sitter spacetime we concluded that the
entropy of the cosmological horizon could not be gravitational entropy.
This leads us to two possible interpretations. Either, horizon entropy
in general is different from gravitational entropy
or there is a thermodynamical factor involved
when we introduce the cosmological constant.
The Schwarzschild and the de~Sitter spacetimes can be viewed as 
two opposite special cases of the SdS spacetime. It might be that the Schwarzschild
spacetime has only gravitational entropy, and the de~Sitter spacetime has only
thermodynamical entropy. This is actually 
what we expected: large thermodynamical entropy in the early universe
and large gravitational entropy around black holes.

In this paper we have approached the problem of gravitational entropy
from a phenomenological point of view instead of developing 
a description from the more  
fundamental properties of gravitation. This is because the classical theory of 
gravitation (general relativity) does not include microscopic states of gravitation.
A working theory of quantum gravity should give a description of gravitational
microstates, and then the gravitational entropy can be found using Boltzmann's 
formula (\ref{eq:boltzmann_entropy}). Until then, phenomenological approaches 
such as this may hopefully give us a hint of what a theory of quantum gravity
will say about the entropy of a gravitational field.

\begin{acknowledgments}
Øystein Rudjord acknowledges support from the Norwegian Research Council 
through the project ``Shedding Light on Dark Energy'' (grant 159637/V30).
\end{acknowledgments}

%%%%%%%%%%%%%%%%%%%%%%%%%%%%%%%%%%%%%%%%%%%%%%%%%%%%%%%%%%%%%%%%%%%%%%%%%%
%%%%%%%%%%%%%%%%%%%%%%%% Appendix %%%%%%%%%%%%%%%%%%%%%%%%%%%%%%%%%%%%%%%%
%%%%%%%%%%%%%%%%%%%%%%%%%%%%%%%%%%%%%%%%%%%%%%%%%%%%%%%%%%%%%%%%%%%%%%%%%%

\appendix*

\section{The relation between the Weyl and the Kretschmann scalars} \label{app}
In this appendix we will show that for static spherically symmetric 
spacetimes,
\[
R^{\alpha\beta\gamma\delta}R_{\alpha\beta\gamma\delta}
\ge C^{\alpha\beta\gamma\delta}C_{\alpha\beta\gamma\delta}.
\]

The Weyl tensor can be expressed by the Riemann tensor as
\[
C_{\alpha\beta\gamma\delta} = R_{\alpha\beta\gamma\delta}-g_{\alpha[\gamma}
R_{\delta]\beta} + g_{\beta[\gamma} R_{\delta]\alpha} + \frac 13 R 
g_{\alpha[\gamma}g_{\delta]\beta}
\]
where $R_{\mu\nu} = R^\lambda_{ \mu\lambda\nu}$ is the Ricci tensor
and $R= R^\lambda_{ \lambda}$ is the Ricci scalar.

From this it is straightforward to show that the Weyl scalar can be
expressed as
\begin{eqnarray}
C^{\alpha\beta\gamma\delta}C_{\alpha\beta\gamma\delta}
&=& R^{\alpha\beta\gamma\delta}R_{\alpha\beta\gamma\delta}
-2R^{\alpha\beta}R_{\alpha\beta} + \frac 13 R^2  \notag\\
&=& R^{\alpha\beta\gamma\delta}R_{\alpha\beta\gamma\delta} + \Delta
\end{eqnarray}
where we have defined the difference $\Delta$ between the
Weyl and Kretschmann scalars
\begin{equation}
\Delta = -2R^{\alpha\beta}R_{\alpha\beta} + \frac 13 R^2.  
\label{eq:deltadef}
\end{equation}
If $\Delta < 0$, then $C^{\alpha\beta\gamma\delta}C_{\alpha\beta\gamma\delta} <
R^{\alpha\beta\gamma\delta}R_{\alpha\beta\gamma\delta}$, and if $\Delta > 0$ then 
$C^{\alpha\beta\gamma\delta}C_{\alpha\beta\gamma\delta} >
R^{\alpha\beta\gamma\delta}R_{\alpha\beta\gamma\delta}$. If $\Delta = 0$ then 
the two scalars are equal.

We begin with the general metric for a static spherically symmetric spacetime:
\[
\diff s^2 =g_{\mu\nu}\diff x^\mu \diff x^\nu=- e^{\nu(r)}c^2 \diff t^2 
+ e^{\lambda(r)}\diff r^2 + r^2 \diff \theta^2 +r^2 \sin^2\theta \diff \phi^2.
\]
We also need the contravariant components of the metric:
\[
g^{\mu\nu} = \mathrm{diag}\left(-e^{-\nu},e^{-\lambda},\frac{1}{r^2}, 
\frac{1}{r^2\sin^2\theta}\right).
\]
The components of the Ricci tensor in this metric is \cite{rindler:relativity}:
\begin{eqnarray}
R_{tt} &=& e^{\nu-\lambda}\left(\frac 12 \nu''-\frac 14 \nu'\lambda'
+\frac 14\nu'^2+\frac{\nu'}{r}\right) \notag \\
R_{rr} &=& -\frac 12 \nu''+\frac 14 \nu'\lambda' -\frac 14 \nu'^2+\frac{\lambda'}{r}
\notag \\
R_{\theta\theta} &=& 1-e^{-\lambda}\left(1+\frac 12 r(\nu'-\lambda')\right) 
\notag \\
R_{\phi\phi} &=& R_{\theta\theta}\sin^2\theta \notag \\
R_{\mu\nu} &=& 0 \quad \mu\ne\nu \notag
\end{eqnarray}
Using these, we find the Ricci scalar:
\begin{eqnarray}
R &=& g^{\mu\nu}R_{\mu\nu} \notag \\
&=& -e^{-\nu}\left[e^{\nu-\lambda}\left(\frac 12 \nu''-\frac 14\nu'\lambda' +
\frac 14 \nu'^2 + \frac{\nu'}{r}\right)\right] \notag \\
&+& e^{-\lambda}\left[-\frac 12 \nu''+\frac 14 \nu'\lambda' -\frac 14 \nu'^2 
+\frac{\lambda'}{r}\right] \notag \\
&+& \frac{2}{r^2} \left[1-e^{-\lambda}\left(1-\frac 12 r(\nu'-\lambda')\right)
\right] \notag \\
&=& -e^{-\lambda}\left(-\frac{2e^{\lambda}}{r^2}+\nu'' -\frac 12 \nu'\lambda'
+\frac 12 \nu'^2 +\frac{2}{r^2}\right) \notag 
\end{eqnarray}
In the expression for $\Delta$ (\ref{eq:delta}) the Ricci scalar appears 
squared, hence
\begin{eqnarray}
R^2 &=& e^{-2\lambda}\left(-\frac{2e^{\lambda}}{r^2}+\nu'' -\frac 12 \nu'\lambda'
+\frac 12 \nu'^2 +\frac{2}{r^2}\right)^2 \notag \\
&=&e^{-2\lambda}\bigg(\nu''^2 -\nu''\nu'\lambda' + \nu''\nu'^2 + \frac{4\nu''}{r^2}
+\frac 14 \nu'^2\lambda'^2 \notag \\
&-& \frac 12 \nu'^3\lambda' - \frac{2\nu'\lambda'}{r^2}
+\frac 14 \nu'^4 + \frac{2\nu'^2}{r^2} + \frac{4}{r^4}\bigg) \notag \\
&+&e^{-\lambda} \left(-\frac{4\nu''}{r^2} + \frac{2\nu''\lambda'}{r^2} - 
\frac{2\nu'^2}{r^2} - \frac{8}{r^4}\right) + \frac{4}{r^4} 
\label{eq:Rscalarsquared}
\end{eqnarray}

Next, we find the ``squared'' Ricci tensor, $R^{\alpha\beta}R_{\alpha\beta}$
\begin{eqnarray}
\begin{split}
R^{\alpha\beta}R_{\alpha\beta} &= g^{\mu\alpha}g^{\nu\beta}R_{\mu\nu}
R_{\alpha\beta}  \\
&=\left(g^{tt}\right)^2\left(R_{tt}\right)^2
+\left(g^{rr}\right)^2\left(R_{rr}\right)^2 \\
&+\left(g^{\theta\theta}\right)^2\left(R_{\theta\theta}\right)^2
+\left(g^{\phi\phi}\right)^2\left(R_{\phi\phi}\right)^2  \\
&= e^{-2\lambda}\bigg(\frac{\nu''^2}{2}-\frac 12 \nu''\nu'\lambda' + \frac 12 \nu''
\nu'^2 
+\frac{\nu''\nu'}{r} \\
&- \frac{\nu''\lambda'}{r}  +\frac 18 \nu'^2\lambda'^2  
-\frac 14 \nu'^3\lambda' - \frac{\nu'^2\lambda'}{r} \\
&+ \frac 18 \nu'^4 
+ \frac{\nu'^3}{2r} + \frac{3\nu'^2}{2r^2} 
+ \frac{\nu'\lambda'^2}{2r} + 
\frac{3\lambda'^2}{2r^3} \\ 
&+ \frac{2}{r^4}  
+ \frac{2\nu'}{r^3} - \frac{2\lambda'}{r^3}
-\frac{\nu'\lambda'}{r^2}\bigg) \\
&- 4e^{-\lambda}\left(\frac{1}{r^4}+
\frac{\nu'-\lambda'}{2r^3}\right) + \frac{2}{r^4} 
\end{split} \label{eq:Rtensorsquared} 
\end{eqnarray}

Inserting (\ref{eq:Rscalarsquared}) and (\ref{eq:Rtensorsquared}) into
(\ref{eq:deltadef}) we find $\Delta$
\begin{equation}
\begin{split}
\Delta &=  -2R^{\alpha\beta}R_{\alpha\beta}+\frac 13 R^2 \\
&= -\frac 16 \bigg(16+4r^4\nu''\nu'^2 - 4r^3\nu''\lambda' +4r^3\nu''\nu' 
+ 8\nu''r^2 e^{\lambda} \\
&- 4r^4\nu''\nu'\lambda' + 4r^4\nu''^2 - 8\nu''r^2 + 4 \nu'^2r^2e^\lambda 
-4\nu'\lambda'r^2e^\lambda \\
&- 32e^\lambda +2r^3\nu'\lambda'^2+r^4\nu''^2 - 2r^4\nu'^3\lambda' 
- 4r^3\nu'^2\lambda' \\
&+ r^4\nu'^4 + 2r^3\nu'^3 +10r^2\lambda'^2 + 16 e^{2\lambda} - 8r\nu'e^{\lambda}
+ 8r\lambda'e^\lambda \\
&+ 8r\nu' -8r\lambda' +8r^2\nu'\lambda' + 6 r^2\nu'^2 \bigg)
/\left(r^4e^{2\lambda}\right).
\end{split}\label{eq:delta}
\end{equation}

If we sort the terms carefully we see that we can extract a factor $(e^\lambda-1)$
from many of them
\begin{equation}
\begin{split}
\Delta &= -\frac{1}{6\left(r^2e^{\lambda}\right)^2}\bigg\{ \left[e^\lambda 
-1\right]
\big[8r^2\nu'' + 4r^2\nu'^2 -4r^2\nu'\lambda'  \\
&- 16 + 16e^{\lambda} - 8r\nu'
+ 8r\lambda'\big]\\
&+ 10r^2\nu'^2 + 4r^2\nu'\lambda' + 4r^4 \nu''\nu'^2 - 4r^3\nu''\lambda' \\
&+4r^3\nu''\nu' - 4r^4\nu'\lambda'\nu'' + 4r^4\nu''^2\\
&+ 2r^3\nu'\lambda'^2 + r^4\nu'^2\lambda'^2 - 2r^4\nu'^3\lambda' \\
&-4r^3 \nu'^2\lambda' + r^4\nu'^4 + 2r^3\nu'^3 + 10 r^2\lambda'^2\bigg\} 
\label{app:delta}
\end{split} 
\end{equation}
and if we look closer at the terms in the parenthesis we notice that for some of 
them we can factor out $(e^\lambda-1)$ again. This, and some more sorting of the 
other terms yields
\begin{eqnarray}
&&\left[e^\lambda -1\right]
\big[8r^2\nu'' + 4r^2\nu'^2 -4r^2\nu'\lambda' \notag \\
&-& 16 + 16e^{\lambda} - 8r\nu'
+ 8r\lambda'\big] \notag \\
&=& \left[e^\lambda -1\right] \big[16\left(e^\lambda -1\right)
+4r\big(2\left\{\lambda'-\nu'\right\}\notag \\
&+& r\left\{2\nu''+\nu'^2-\nu'\lambda'\right\}
\big)\big] \notag \\
&=& 16\left[e^\lambda -1\right]^2 + \left[e^\lambda -1\right]
4r\big[2\left(\lambda'-\nu'\right) \notag \\
&+&r\left(2\nu''+\nu'^2-\nu'\lambda'\right)
\big] \notag 
\end{eqnarray}
We see that we almost have the square of a sum. Adding and subtracting the same 
term to complete the square brings us to
\begin{eqnarray}
&&16\left[e^\lambda -1\right]^2 + \left[e^\lambda -1\right]
4r\big[2\left(\lambda'-\nu'\right)\notag \\
&&+r\left(2\nu''+\nu'^2-\nu'\lambda'\right)
\big] \notag \\
&=&16\left[e^\lambda -1\right]^2 + \left[e^\lambda -1\right]
4r\big[2\left(\lambda'-\nu'\right)\notag \\
&&+r\left(2\nu''+\nu'^2-\nu'\lambda'\right)
\big] \notag \\
&&+ \frac 14 r^2 \left[2\left(\lambda'-\nu'\right) 
+ r\left(2\nu''+\nu'^2-\lambda'\nu'\right) \right]^2 \notag \\
&&- \frac 14 r^2 \left[2\left(\lambda'-\nu'\right) 
+ r\left(2\nu''+\nu'^2-\lambda'\nu'\right) \right]^2 \notag \\
&=& \left[ 4\left(e^\lambda -1\right)+\frac 12 r\left(2\left\{\lambda'-\nu'\right\}
+r\left\{2\nu''+\nu'^2-\nu'\lambda'\right\}\right)\right]^2 \notag \\
&&- \frac 14 r^2 \left[2\left(\lambda'-\nu'\right) 
+ r\left(2\nu''+\nu'^2-\lambda'\nu'\right) \right]^2 \notag
\end{eqnarray}
Inserting this into the expression for $\Delta$ (\ref{app:delta}) and sorting the 
other terms a bit gives us
\begin{equation}
\begin{split}
\Delta &= -\frac{1}{6\left(r^2e^{\lambda}\right)^2}\Bigg\{
\bigg[ 4\left(e^\lambda -1\right) \\ 
&+\frac 12 r\Big(2\left\{\lambda'-\nu'\right\} 
+r\left\{2\nu''+\nu'^2-\nu'\lambda'\right\}\Big)\bigg]^2  \\
&- \frac 14 r^2 \left[2\left(\lambda'-\nu'\right) 
+ r\left(2\nu''+\nu'^2-\lambda'\nu'\right) \right]^2 \\
&+ r^4\left[4\nu''^2 + 4\nu''\nu'^2 -4\nu''\nu'\lambda' + \nu'^4 - 2\nu'^3\lambda'
+ \nu'^2\lambda'^2 \right]\\
&-2 r^3\left[2\nu''\lambda'
+2\nu'^2\lambda' - \nu'\lambda'^2 - 2\nu''\nu' - \nu'^3 \right] \\
&+2r^2\left[5\lambda'^2 - 2\lambda'\nu' + 5\nu'^2\right] \Bigg\}
\end{split} \label{app:delta2}
\end{equation}
We split the fifth term and write it as a sum of two squares
\begin{eqnarray}
&&2r^2\left[5\lambda'^2 - 2\lambda'\nu' + 5\nu'^2\right] \notag \\
&=& 4r^2\left[\lambda'^2-2\lambda'\nu'+\nu'^2\right] +
6r^2\left[\lambda'^2 + 2\lambda'\nu' + \nu'^2 \right] \notag \\
&=& 4r^2\left[\lambda'-\nu'\right]^2 
+ 6r^2\left[\lambda'+\nu'\right]^2  \label{app:term5}
\end{eqnarray}
and if we look closer we also see that the third term of (\ref{app:delta2}) is a 
square
\begin{eqnarray}
&&r^4\left[4\nu''^2 + 4\nu''\nu'^2 -4\nu''\nu'\lambda' + \nu'^4 - 2\nu'^3\lambda'
+ \nu'^2\lambda'^2 \right] \notag \\
&=& r^4\left[2\nu''+\nu'^2-\nu'\lambda'\right]^2
\label{app:term3}
\end{eqnarray}
and the fourth term can be written as a product
\begin{equation}
\begin{split}
-2r^3&\left[2\nu''\lambda'
+2\nu'^2\lambda' - \nu'\lambda'^2 - 2\nu''\nu' - \nu'^3 \right]\\
&= -2r^3\left[\lambda'-\nu'\right]\left[2\nu''+\nu'^2-\nu'\lambda'\right]
\end{split}\label{app:term4}
\end{equation}
Inserting (\ref{app:term5}), (\ref{app:term3}), (\ref{app:term4}) into 
(\ref{app:delta2}) and writing out the second term we get
\begin{eqnarray}
\Delta &=& -\frac{1}{6\left(r^2e^{\lambda}\right)^2}\bigg\{
\bigg[ 4\left(e^\lambda -1\right)\notag \\
&+&\frac 12 r\Big(2\left\{\lambda'-\nu'\right\}
+r\left\{2\nu''+\nu'^2-\nu'\lambda'\right\}\Big)\bigg]^2  \notag \\
&-& r^2 \left[\lambda'-\nu'\right]^2 -r^3\left[\lambda'-\nu'\right]
\left[2\nu''+\nu'^2-\lambda'\nu'\right] \notag \\
&-& \frac 14 r^4 \left[2\nu''+\nu'^2-\lambda'\nu'\right]^2 
+ r^4\left[2\nu''+\nu'^2-\nu'\lambda'\right]^2 \notag \\
&-& 2r^3\left[\lambda'-\nu'\right]\left[2\nu''+\nu'^2-\nu'\lambda'\right]
+4r^2\left[\lambda'-\nu'\right]^2 \notag \\
&+& 6r^2\left[\lambda'+\nu'\right]^2  \bigg\} \notag \\
&=& -\frac{1}{6\left(r^2e^{\lambda}\right)^2}\bigg\{
\bigg[ 4\left(e^\lambda -1\right)\notag \\
&+&\frac 12 r\Big(2\left\{\lambda'-\nu'\right\}
+r\left\{2\nu''+\nu'^2-\nu'\lambda'\right\}\Big)\bigg]^2  \notag \\
&+& 3r^2\left[\lambda'-\nu'\right]^2 
- 3r^3 \left[\lambda'-\nu'\right] \left[2\nu''+\nu'^2-\lambda'\nu'\right] \notag \\
&+& \frac 34 r^4\left[2\nu''+\nu'^2-\lambda'\nu'\right]^2 
+ 6r^2\left[\lambda'+\nu'\right]^2  \bigg\} \notag
\end{eqnarray}
Once again, we recognize a square, and rewrite the second, third and fourth term
to get the final expression
\begin{eqnarray}
\begin{split}
\Delta &= -\frac{1}{6\left(r^2e^{\lambda}\right)^2}\bigg\{
\bigg[ 4\left(e^\lambda -1\right) \\
&+\frac 12 r\Big(2\left\{\lambda'-\nu'\right\} 
+r\left\{2\nu''+\nu'^2-\nu'\lambda'\right\}\Big)\bigg]^2  \\
&+\frac 34 r^2\big[r \left(2\nu''+\nu'^2-\lambda'\nu'\right) \\
&-2\left(\lambda'-\nu'\right)\big]^2
+ 6r^2\left[\lambda'+\nu'\right]^2 \bigg\}
\end{split}\label{app:delta3}
\end{eqnarray}
containing only square terms. Thus $\Delta \le 0$, and consequently
\begin{equation}
R^{\alpha\beta\gamma\delta}R_{\alpha\beta\gamma\delta}
\ge C^{\alpha\beta\gamma\delta}C_{\alpha\beta\gamma\delta}.
\end{equation}

\bibliography{wcc_paper}
\end{document}